# Bayesian Identification of Elastic Constants in Multi-Directional Laminate from Moiré Interferometry Displacement Fields.

C. Gogu[1], W. Yin[2], R. Haftka[2], P. Ifju[2], J. Molimard[3], R. Le Riche[4], and A. Vautrin[5†]

[1] Université de Toulouse; INSA, UPS, Mines Albi, ISAE; ICA (Institut Clément Ader); 118, route de Narbonne, F-31062 Toulouse, France
[2] University of Florida, Mechanical and Aerospace Engineering Department, PO Box 116250, Gainesville, FL, 32611, USA
[3] Ecole Nationale Supérieure des Mines, CIS-EMSE, CNRS :UMR5146, LCG, F-42023 Saint Etienne, France
[4] Ecole Nationale Supérieure des Mines, FAYOL-EMSE, CNRS :UMR5146, LCG, F-42023 Saint Etienne, France
[5] Ecole Nationale Supérieure des Mines, SMS-EMSE, CNRS :UMR5146, LCG, F-42023 Saint Etienne, France

**Abstract.** The ply elastic constants needed for classical lamination theory analysis of multi-directional laminates may differ from those obtained from unidirectional laminates because of three dimensional effects. In addition, the unidirectional laminates may not be available for testing. In such cases, full-field displacement measurements offer the potential of identifying several material properties simultaneously. For that, it is desirable to create complex displacement fields that are strongly influenced by all the elastic constants. In this work, we explore the potential of using a laminated plate with an open-hole under traction loading to achieve that and identify all four ply elastic constants ($E_1$, $E_2$, $\nu_{12}$, $G_{12}$) at once. However, the accuracy of the identified properties may not be as good as properties measured from individual tests due to the complexity of the experiment, the relative insensitivity of the measured quantities to some of the properties and the various possible sources of uncertainty. It is thus important to quantify the uncertainty (or confidence) with which these properties are identified. Here, Bayesian identification is used for this purpose, because it can readily model all the uncertainties in the analysis and measurements, and because it provides the full coupled probability distribution of the identified material properties. In addition, it offers the potential to combine properties identified based on substantially different experiments. The full-field measurement is obtained by moiré interferometry. For computational efficiency the Bayesian approach was applied to a proper orthogonal decomposition (POD) of the displacement fields. The analysis showed that the four orthotropic elastic constants are determined with quite different confidence levels as well as with significant correlation. Comparison with manufacturing specifications showed substantial difference in one constant, and this conclusion agreed with earlier measurement of that constant by a traditional four-point bending test. It is possible that the POD approach did not take full advantage of the copious data provided by the full field measurements, and for that reason that data is provided for others to use (as on line material attached to the article).

# 1 Introduction

Orthotropic elastic constants of composite materials are usually identified by series of tests on unidirectional laminate coupons. This provides good accuracy, however, it is not always possible or appropriate. First, one may want to obtain the elastic constants of laminates made some time ago, for reverse engineering, to determine the effect of aging, or to determine whether the elastic constants do not change somewhat due to the interaction between plies of different fiber angles combined in a single laminate. In this situation, it has been suggested that a full-field measurement of the response might furnish sufficient data to identify all the properties from a single test if the strain field in the laminate is complex enough.

A tensile test on an elastically orthotropic plate with a hole can create such a complex displacement field and has already been used in the past for identifying the four orthotropic elastic constants of a laminate [1,2]. This is also the experiment that we will use in the present paper. The objective of the paper is to find how accurately we can find the elastic constants from such a single test.

After the experiment, we need to decide which method to use for the identification itself. Multiple approaches exist such as least squares based finite element model updating, the constitutive equation gap method, the virtual fields method, the equilibrium gap method, the reciprocity gap method and we refer the reader to the review by Avril et al. [3] for an overview of each of these methods in the context of full field based identification.

While the above methods were designed to provide numerical values for the properties to be identified, characterizing the uncertainty in the identified properties is sometimes crude (only variance estimates but no covariances) and in any case not systematic. This is important however, since different material properties determined based on a single test are not identified with the same confidence. Typically the highest uncertainty is associated with respect to properties to which the experiment is the most insensitive. In addition, the uncertainty in different properties can be strongly correlated, so that obtaining only variance estimates may be misleading.

One of the challenges in the identification of multiple material properties from a complex experiment resides in handling different sources of uncertainty in the experiment and the modelling of the experiment for estimating the resulting uncertainty in the identified properties. A possible approach for doing this is the Bayesian method [4,5]. This method was introduced in the late 1970s in the context of identification [6] and has been applied since to different problems, notably identification of elastic constants from plate vibration experiments [7,8]. The applications of the method to these classical point-wise tests involved only a small number of measurements (typically ten natural frequencies in the previously cited vibration test), which facilitated the application of the Bayesian approach.

In the present article we adopt a Bayesian framework for identifying the orthotropic elastic constants of a composite material from an open-hole tensile test, on which we measure the displacement fields. Several authors carried out identifications based on such measurements within a least squares model updating framework [9,10]. We apply here the Bayesian identification approach to moiré interferometry displacement measurements with the aim of identifying the four ply elastic constants of a composite laminate.

The rest of the paper is organized as follows. In Section 2 we give an overview of the identification problem from an open-hole tensile test as well as the moiré interferometry experiment that was carried out. Section 3 provides an overview of the numerical modelling allowing efficient treatment of the identification problem. In section 4 we provide the Bayesian identification formulation, obtained results and their discussion. We give concluding remarks in Section 5.

## 2 Open hole tensile test

### 2.1 Experiment

In this paper we identify the orthotropic ply-elastic constants from full field displacement measurements on an open-hole plate. The plate is a laminate made from a graphite/epoxy prepreg (Toray® T800/3631) with a stacking sequence of [45,-45,0]$_s$. Prior information on the properties that we seek was available from the manufacturer and from previous experiments. The manufacturer's specifications are given in Table 1 together with the properties obtained by Noh [11]. Noh obtained the material properties based on a four-point bending test at the University of Florida on a laminate made from the exact same prepreg roll that we used for the specimen of this study.

The hole in the plate was machined using a special drill designed specifically for composites. Special care was taken to avoid delaminations around the hole especially on the drill exit surface, since these delaminations may change the displacement field around the hole. Only those specimens without visible delamination around the hole were kept for testing. The absence of significant delaminations was also confirmed after the measurements by the fact that the measured displacement fields did not show any significant deviations around the hole from the expected fields.

Table 1. Manufacturer's specifications and properties found by Noh [11] based on a four points bending test.

| Parameter | $E_1$ (GPa) | $E_2$ (GPa) | $\nu_{12}$ | $G_{12}$ (GPa) |
|---|---|---|---|---|
| Manufacturer's specifications | 162 | 7.58 | 0.34 | 4.41 |
| Noh's values [11] | 144 | 7.99 | 0.34 | 7.78 |

In the present study we seek to identify the ply-properties and their uncertainties from a tensile test on a laminate having the dimensions given in Figure 1. The total thickness *h* is 0.78 mm and the applied tensile force is 700 N. The *U* and *V* displacement fields are defined as being in the 1 and 2 direction, respectively. The full field measurements are taken on a square area 24.3 x 24.3 mm$^2$ around the center of the hole.

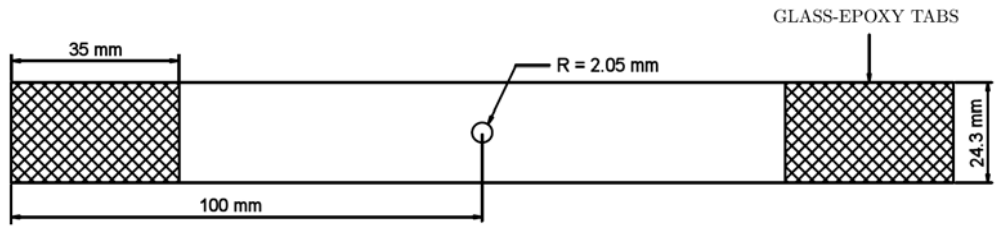

Figure 1. Specimen geometry. The specimen material is graphite/epoxy and the stacking sequence [45,-45,0]$_s$. The tensile force is 700 N.

The experimental setup utilized is shown in Figure 2. The testing machine was an MTI-30K. Rotations of the grips holding the specimen were allowed by using a lubricated ball bearing for the bottom grip and two lubricated shafts for the top grip. This allowed to reduce parasitic bending during the tension test. An ESM Technologies PEMI II 2020-X moiré interferometer using a Pulnix TM-1040 digital camera were utilized.

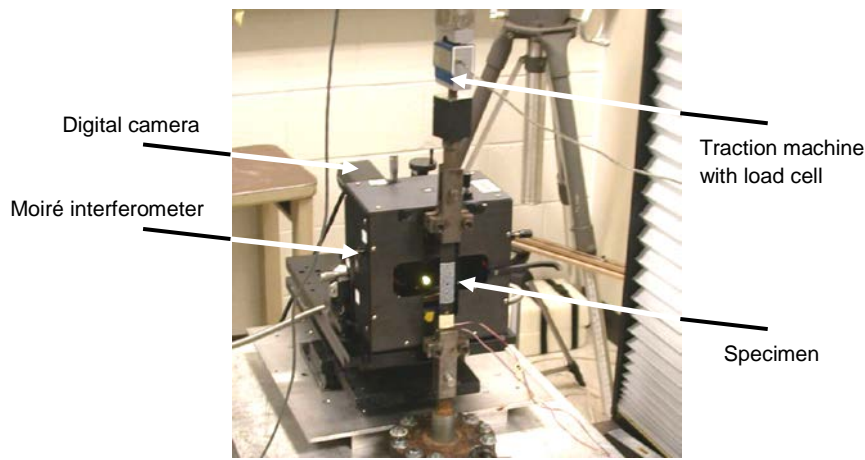

Figure 2. Experimental setup for the open-hole tension test.

Moiré interferometry (Post et al. [12]) is a measurement technique using the fringe patterns obtained by optical interference off a diffraction grating in order to obtain full-field displacement or strain maps. Among its main advantages are high signal to noise ratio, excellent spatial resolution and insensitivity to rigid body rotations [13]. The displacement resolution, obtained by repeatability tests, can be as low as 4 nm. Past applications of moiré interferometry include the mapping of displacements of a tooth [14] and characterization of advanced fabric composites [15]. Additional applications are also provided by Post et al. [12].

The schematic of a four-beam moiré interferometry setup used for the present experiment is given in Figure 3. It uses four collimated light beams, thus providing both the horizontal and vertical displacement fields. The interference is obtained by choosing the angle α such that it corresponds to the first order diffraction angle. In this case two opposite beams will be diffracted/reflected of the Moiré grating placed on the specimen. Since the camera is positioned at the angle of the first diffraction order it will capture the interference fringe pattern with the highest intensity.

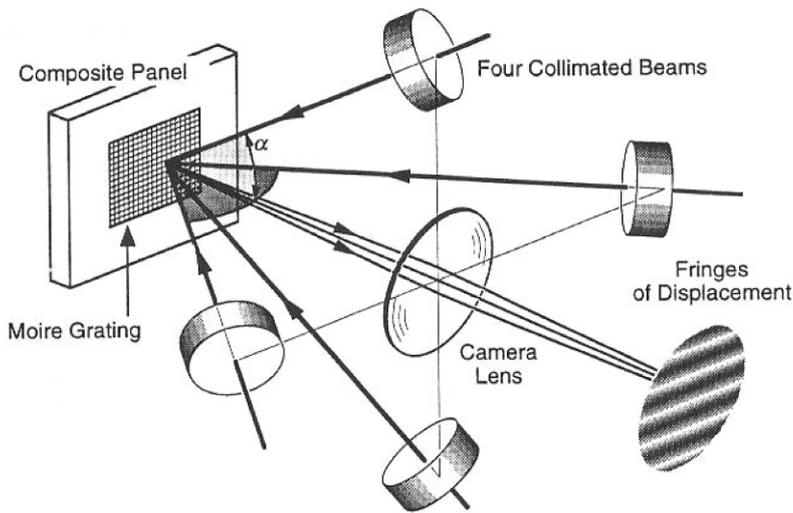

Figure 3. Schematic of a moiré interferometry setup.

The fringe patterns that result from the interference of two of the beams can be described by either intensity or phase information. While intensity methods have been developed first, a major issue limiting their accuracy resides in the determination of the exact maximum intensity locations. To address this issue, methods based on phase information were developed, such as phase shifting moiré. All these methods use a carrier fringe pattern or a phase ramp in order to extract the phase φ, due to the fact that the cosine function is not bijective. Using a phase shift λ, the intensity I can then be expressed as shown in Equation 1.

$$I(x, y) = I_{backlight}(x, y) + I_{mod}(x, y)\cos[\varphi(x, y) + n\lambda] \qquad n = 1...N \qquad (1)$$

Obtaining N fringe patterns (typically N=4) shifted by the imposed phase shift allows to calculate the phase φ(x,y). The displacement fields are then determined as follows:

$$U(x, y) = \frac{\Delta \varphi_x(x, y)}{2\pi f_s} \qquad (2)$$

$$V(x, y) = \frac{\Delta \varphi_y(x, y)}{2\pi f_s} \qquad (3)$$

where Δφ is the phase difference between the initial and the final loading step and $f_s$ is the spatial frequency of the reference grating (2400 lines/mm in our case).

An automated phase extraction procedure was developed under Matlab by Yin [16] at the Experimental Stress Analysis Laboratory at the University of Florida. This toolbox, which will be used here, carries out the phase extraction and unwrapping from the four phase shifted moiré fringe patterns. It then provides the corresponding displacement fields.

The displacement fields obtained are provided as data files attached to the online version of this article (see Appendix 4) and are also graphically illustrated in Figure 4. Note that no filtering whatsoever was used during the extraction algorithm. These two displacement fields serve as the measurements for the present identification problem.

Typical sources of uncertainty affecting the displacement fields are noise, the phase extraction procedure, imperfect centering of the hole on the specimen or misalignment of the grips, which can create bending .

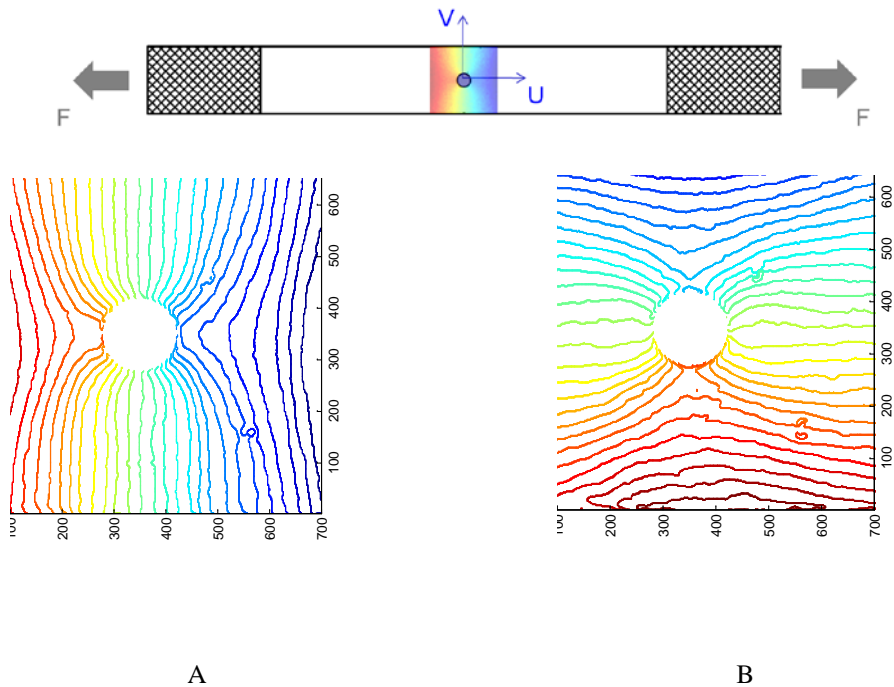

A                                            B

Figure 4. Displacement fields (in µm) obtained from the fringe patterns in: A) The 1-direction (*U* field). B) The 2-direction (*V* field).

## 2.2 Modelling and problem statement

In order to identify the ply-elastic constants, $E_1$, $E_2$, $v_{12}$, $G_{12}$, we need a model relating these to the displacement fields. Unfortunately there are no exact analytical solutions for the problem of an orthotropic plate. Instead we chose a finite element model that will be used for the Bayesian model updating.

The plate is modeled using the Abaqus® finite element software. A total of 8020 S4R elements (general purpose, four nodes per element, reduced integration) were used. Boundary conditions were imposed by prescribed forces. The finite element mesh in the area of interest is shown in Figure 5 and the measurement area highlighted in red. Note that Figure 5 does not include the entire mesh. Since the whole plate is modeled in Abaqus there is a transition using triangular elements towards a larger mesh at the grip edges of the plate where the stresses are relatively uniform compared to the area around the hole. The irregularity of this transition did not have any significant effect on the displacements in the area of interest.

A finite element mesh convergence study was carried out to assess the quality of the model, and it was found that with the present mesh the discretization error in the area of interest was of the order of $6 \times 10^{-4}$ % of the average absolute value of the field, which was considered acceptable.

Note that the model uses shell elements for computing the mid-plane displacement fields after homogenization of the six-plies composite into a single orthotropic layer. With in-plane loading we considered that the computations using the homogenized model lead to acceptable results and that we can consider that the mid-plane displacement field is the same as the surface displacement field. The absence of any significant out of plane bending induced by the experimental setup was verified on some specimen by the use of stereo digital image correlation on the other side of the specimen.

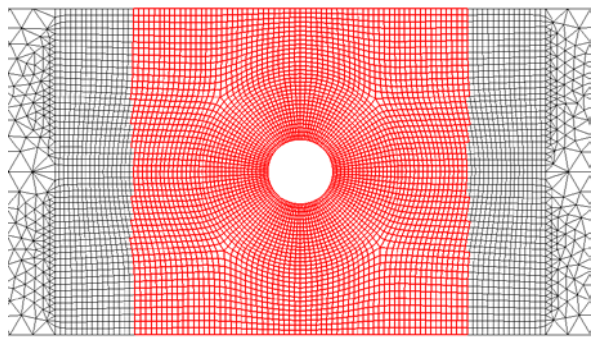

Figure 5. Finite element mesh. The measurement area is highlighted in red.

During the identification process we vary a certain number of model parameters such as elastic constants or plate dimensions and obtain each time the corresponding full fields. In model updating frameworks one seeks to match the predictions with the experimental fields either in a deterministic way (least squares) or in a probabilistic way (Bayesian), as we do here.

The field is described here by the displacement values at the 4569 nodes within the reference area (see highlighted area in Figure 5). Note that the experimental fields contain much more information, since we obtain 490,000 measurement points (pixels) per field. The values of the experimental fields at the node positions are calculated by linear interpolation.

If a field calculation needs to be used within the Bayesian framework where correlation between the measurements is required, it is not practical to describe the fields by their value at each point (at the 4569 nodes here). This is essentially because thousands-dimensional probability density functions required to describe the correlation between the different measurement points are outside the realm of what statistical methods can currently handle with reasonable computational resources.

Furthermore the model evaluation needs to be repeated millions of times during the identification, a problem exacerbated by the need for statistical sampling. Using a finite element model directly is not computationally feasible in this case.

To address these problems we use the proper orthogonal decomposition method for dimensionality reduction and response surface methodology for cost reduction. These are described in the following section.

## 3 Numerical modelling

### 3.1 Proper orthogonal decomposition of the full-fields

The aim of the proper orthogonal decomposition (POD) method is to construct an optimal, reduced dimensional basis for the representation of simulation samples (that are called snapshots in POD terminology).

Let us consider $\boldsymbol{U}^i \in \mathbb{R}^n$, which is the vector representation of a field (e.g. displacement field). Note that $n$ is usually several thousands. We seek, based on $N$ sample vectors $\{\boldsymbol{U}^i\}_{i=1..N}$, a reduced dimensional representation of the fields' variations with some input parameters.

For the open-hole plate identification problem we are interested in accounting for variations of the following parameters: ply elastic constants $E_1$, $E_2$, $v_{12}$, $G_{12}$ and ply thickness $t$. We are looking at variations of the homogenized ply-properties and thickness here and not at spatial variations within the plate. Accounting for variations in the elastic constants is needed as usual for the identification procedure. We added here the ply thickness as an important source of uncertainty. We assumed here that we are interested in variations of the parameters $E_1$, $E_2$, $v_{12}$, $G_{12}$ and $t$ within the bounds given in Table 2.

Table 2. Bounds on the input parameters of interest (for a graphite/epoxy composite material).

| Parameter | $E_1$ (GPa) | $E_2$ (GPa) | $v_{12}$ | $G_{12}$ (GPa) | $t$ (mm) |
| --- | --- | --- | --- | --- | --- |
| Lower bound | 126 | 7 | 0.189 | 3.5 | 0.12 |
| Upper bound | 234 | 13 | 0.351 | 6.5 | 0.18 |

We obtained the snapshots required for the POD approach by sampling 200 points within the bounds of Table 2. These bounds were chosen such that the manufacturer's specifications roughly lie in the middle. We chose relatively wide bounds reflecting the fact that we did not want to restrict the area in which the identified properties are sought. The points are sampled by Latin hypercube, which consists in obtaining the 200 sample points by dividing the range of each parameter into 200 sections of equal marginal probability 1/200 and sampling once from each section. Latin hypercube sampling typically ensures that the points are reasonably well distributed in the entire space.

At each of the 200 sampled points we then perform a finite element analysis, which gives the corresponding horizontal and vertical displacement fields $U$ and $V$ respectively. Based on these samples we construct the POD basis using the singular value decomposition as detailed in Appendix 1. In the obtained POD basis we can express approximately any displacement field obtained for parameter values within the bounds of Table 2 as a linear combination of the POD basis vectors:

$$\widetilde{U}^i = \sum_{k=1}^{K} \alpha_{i,k} \Phi_k = \sum_{k=1}^{K} \left\langle U^i, \Phi_k \right\rangle \Phi_k \tag{4}$$

where $\widetilde{U}^i$ is the POD approximation of the displacement field $U^i$ (which can be either a U displacement field (in the 1-direction) or a V displacement field (in the 2-direction); $\left\{ \Phi_k \right\}_{k=1..K}$ are the basis vectors (also called POD modes) of the POD orthogonal basis and $\alpha_{i,k}$ are the coefficients of the displacement field in this basis, which can be obtained by the orthogonal projection of field $U^i$ onto the basis vectors.

Once the POD modes (basis vectors) were determined by singular value decomposition, we still need to find an appropriate truncation order $K$ for the reduced dimensional approximations of the fields (see Equation 4).

In order to choose the truncation order we first use a typical error criterion for POD the POD method based on the norm of the residuals (see Eq. 10 in Appendix 1 for details). Table 3 provides this truncation error criterion for various truncation orders. An additional global error criterion is provided in Appendix 1.

Table 3. Error norm truncation criterion ($\varepsilon$ is defined in Equation 10 of Appendix 1).

| K | 2 | 3 | 4 | 5 |
|---|---|---|---|---|
| $\varepsilon$ for U fields | $2.439 \times 10^{-7}$ | $4.701 \times 10^{-9}$ | $7.280 \times 10^{-11}$ | $1.211 \times 10^{-11}$ |
| $\varepsilon$ for V fields | $1.054 \times 10^{-6}$ | $2.900 \times 10^{-9}$ | $4.136 \times 10^{-10}$ | $3.517 \times 10^{-11}$ |

Since the fields will be used for identification, not only the accuracy of the fields is important but also the accuracy of the derivatives of the field with respect to the ply-elastic constants. This was verified and we found that four POD coefficients for each field are also sufficient for representing the derivates accurately enough. For details on this verification the reader is referred to [19] (Chapter 6).

On a final note, the identification procedure will use the POD projection of the displacement fields, which filters out some information present in the initial fields. This can have both positive and negative effects. Obvious negative effects are that the identification procedure will not be able to account for any information that was filtered out and that might have been useful to the identification or the propagation of uncertainties. On the other hand if the information filtered out is mainly related to the analysis tools used (e.g. phase extraction algorithm) or to numerical or experimental noise it can be useful to filter these elements out since they do not have physical meaning in relation to the material properties. For an initial investigation of the errors modes left out by the use of the POD procedure the reader is referred to [18] (Chapter 7).

**3.2 Response surface approximations**

Even though we reduced the dimensionality of the full-field using the POD decomposition, the calculation of the POD coefficients is still based on finite element results. Since the Bayesian identification procedure (cf. next section) needs millions of evaluations, we construct computationally cheap approximations of the POD coefficients, $\alpha_k$, as functions of the four elastic constants to be identified and the thickness of the plate.

For this purpose we use response surface approximations (RSA). In particular we used polynomial response surfaces (PRS), which fit the simulation at sample points with a polynomial so as to minimize the square difference between the simulations and the prediction of the PRS. The accuracy of the approximation can then be estimated using indicators such as root mean square (RMS) error or cross validation error. For more details on RSA techniques the reader can refer to [19].

Here we fitted cubic polynomial response surface approximations for each POD coefficient, of the form $\alpha_k$=PRS($E_1$, $E_2$, $\nu_{12}$, $G_{12}$, $h$) to the same 200 samples that were used in the previous section to construct the POD basis. These 200 points were sampled using Latin hypercube within the bounds given in Table 2.

Several error measures were used to assess the accuracy of the obtained response surface approximations and these are provided in Appendix 2. These error measures showed that the approximation error was negligible compared to the other sources of uncertainty.

## 4 Bayesian identification

### 4.1 Bayesian formulation

We identify the joint probability distribution of the elastic constants $E_1$, $E_2$, $\nu_{12}$, $G_{12}$ given the measured displacement fields using a Bayesian formulation. Denoting by $f$ the probability density functions (PDF), then the PDF that we seek, also called posterior PDF, is given by Bayes' formula:

$$f_{E/\alpha=\alpha^{measure}}(\boldsymbol{E}) = \frac{1}{K} f_{\alpha/E}(\boldsymbol{\alpha}^{measure}) \cdot f_E^{prior}(\boldsymbol{E}) \tag{5}$$

where $\boldsymbol{E} = \{E_1, E_2, \nu_{12}, G_{12}\}$ is the four dimensional random variable of the ply-elastic constants. $\boldsymbol{\alpha} = \{\alpha_1^U, ..., \alpha_4^U, \alpha_1^V...\alpha_4^V\}$ is the eight dimensional random variable of the POD coefficients of the $U$ and $V$ field; $\boldsymbol{\alpha}^{measure} = \{\alpha_1^{U,measure}, ..., \alpha_4^{U,measure}, \alpha_1^{V,measure}...\alpha_4^{V,measure}\}$ is the vector of the eight "measured" POD coefficients, obtained by projecting the measured full fields onto the POD basis.

Equation 5 provides the joint probability density function (PDF) of the four elastic constants given the coefficients $\boldsymbol{\alpha}^{measure}$. This PDF, also called posterior PDF and denoted $f_{E/\alpha=\alpha^{measure}}(\boldsymbol{E})$, is equal to a normalizing constant $K$ times the likelihood function of the elastic constants $\boldsymbol{E}$ given the coefficients $\boldsymbol{\alpha}^{measure}$ times the prior distribution of the elastic constants $\boldsymbol{E}$.

The prior distribution of $\boldsymbol{E}$ reflects the prior knowledge we have on the elastic constants. In our case this prior knowledge stems from two sources: the manufacturer's specifications and a four point bending test carried out by Noh on a laminate made from the same prepreg roll that we used (cf. Table 1 for the corresponding values). Since we do not have any information about the accuracy of the values in either source we chose to use a relatively wide prior distribution with 10% standard deviation and no correlation. The mean value of the prior could be chosen to be either Toray's® specifications or Noh's values or a mix of the two (e.g. average). We showed in [18] (Chapter 7) that because of the wide prior (10% standard deviation) there is little influence of the mean values chosen on the posterior PDF. We chose to present here the results based on a prior based on the manufacturer's specifications (see Table 4 for the parameters of the prior used) but the interested reader can refer to [18] for the results based on Noh's prior. The prior distribution was truncated at the bounds given in Table 5, which were chosen in an iterative way such as to verify following aspects. First the bounds were chosen such that eventually the mean of the posterior PDF is approximately in the center of the bounds. Second, the range between the lower and upper truncation bounds was chosen as about four standard deviations of the posterior PDF for $E_2$ and $v_{12}$, for which the Toray specifications and Noh's values agree well. For $E_1$ and $G_{12}$ for which the agreement is worse we chose a range of about eight standard deviations of the posterior PDF between the truncation bounds. Note that it may seem strange that the truncation bound for $E_1$ lies at the mean value of the prior, but this is due to the fact that the mean of the posterior was found to be quite far away from the manufacturer's specification (mean of the prior) as will be discussed later in section 4.2.

Table 4. Normal uncorrelated prior distribution of the material properties for a graphite/epoxy composite material.

| Parameter | $E_1$ (GPa) | $E_2$ (GPa) | $v_{12}$ | $G_{12}$ (GPa) |
|---|---|---|---|---|
| Mean value | 162 | 7.58 | 0.34 | 4.41 |
| Standard deviation | 16 | 0.75 | 0.03 | 0.5 |

Table 5. Truncation bounds on the prior distribution of the material properties

| Parameter | $E_1$ (GPa) | $E_2$ (GPa) | $v_{12}$ | $G_{12}$ (GPa) |
|---|---|---|---|---|
| Lower truncation bound | 118 | 6 | 0.26 | 4.25 |
| Upper truncation bound | 162 | 9.5 | 0.36 | 5.75 |

Other than the prior, the other term on the right hand side of Equation 5 is the likelihood function of the elastic constants given the POD coefficients $\alpha^{measure}$. This function provides an estimate of the likelihood of different $E$ values given the test results.

The uncertainty in the POD coefficients can have several causes, which are detailed next. An important cause is measurement error. Full-field measurements are usually noisy, and the measured field can be decomposed into a signal component and a white noise component. A Gaussian white noise on the full-fields leads also to Gaussian distributions on the POD coefficients, with zero mean and the same standard deviation as the noise on the fields showed (cf. [18] (Chapter 7) for the proof). Note that this does not mean that there is no filtering effect through the use of the POD coefficients; while the standard deviations are the same the resulting fields will be different since the noise does not act on the same quantities (POD coefficients versus displacement values).

Another uncertainty in the identification process is due to uncertainty in the other input parameters of the plate model such as the thickness, misalignment of the center of the hole or misalignment of the loading direction. The thickness of the plate $h$ was assumed to be distributed normally with a mean value of 0.78 mm (the prescribed specimen thickness) and a standard deviation of 0.005 mm (the typical accuracy of a microcaliper). Note that this uncertainty accounts for the lack of knowledge on the actual spatially averaged thickness of the specimen. Since boundary conditions are force-prescribed, the assumed uncertainty of about 0.6% in the thickness leads to an uncertainty of opposite sign in the moduli. This is taken into account through the likelihood function in the Bayesian approach. However, since the various uncertainties add up as their squares, the thickness uncertainty appears to contribute only a small part of the total uncertainty.

Finally, alignment uncertainty as well as other sources of modeling uncertainty for the calculation of the likelihood function were considered indirectly, with somewhat decreased fidelity through a generic uncertainty term on the POD coefficients that had zero mean and a standard deviation of 0.4% of the mean value of the POD coefficients.

In order to calculate the posterior probability density function of Eq. 5 a numerical procedure previously developed by the authors was used. Details on this approach are provided in Appendix 3. This procedure was first tested on simulated full field measurements, where good agreement between the true values and most likely identified values of the properties was found. For details on the identification on the simulated experiment the reader is referred to [18] (Chapter 7).

**4.2 Identification results and discussion**

The Bayesian framework does not identify a single value for each of the four ply-elastic constants but a probability distribution function characterizing the properties as well as the uncertainties with which these are obtained for the specific specimen and the specific experiment and numerical models used.

Applying the Bayesian procedure to the experimental displacement fields described in Section 2 leads to the four-dimensional approximately normal joint probability distribution having the mean values, coefficient of variations (i.e. standard deviation over mean value) and correlation matrix given in Tables 6 and 7. By doing the identification with increasingly larger grid sizes for the support of the PDF the mean values were found to be converged to about one percent, the standard deviations were found to be converged to about 8% while the correlations coefficients were found to be converged to about 10%.

Table 6. Mean values and coefficients of variation of the identified posterior distribution based on the moiré interferometry full-fields from an open-hole tensile test.

| Parameter | $E_1$ (GPa) | $E_2$ (GPa) | $v_{12}$ | $G_{12}$ (GPa) |
|---|---|---|---|---|
| Mean value | 140 | 7.48 | 0.33 | 5.02 |
| COV (%) | 3.3 | 9.5 | 10.3 | 4.3 |

Table 7. Correlation matrix (symmetric) of the identified posterior distribution based on the moiré interferometry full fields from an open-hole tensile test.

|  | $E_1$ | $E_2$ | $v_{12}$ | $G_{12}$ |
|---|---|---|---|---|
| $E_1$ | 1 | 0.023 | -0.043 | 0.51 |
| $E_2$ | - | 1 | -0.005 | -0.17 |
| $v_{12}$ | - | - | 1 | 0.24 |
| $G_{12}$ | - | - | - | 1 |

We note first that the coefficients of variation with which the properties are identified vary greatly from one property to another. While the longitudinal Young's modulus $E_1$ of the ply is identified most accurately, the Poisson's ratio $v_{12}$ of the ply is identified with the highest uncertainty. This trend has been noted before in the composites community, since the measured quantities (displacements here) are typically less sensitive to ply's Poisson's ratio compared to the other three properties when all of them are being sought simultaneously from a complex experiment on a multi-ply laminate. For example identifying the orthotropic constants based on vibration experiments also led the Poisson's ratio to be identified the least accurately [23]. Back to the present experiment we also note that $E_2$ is identified here with a higher uncertainty than $G_{12}$. This is due to the stacking sequence [45,-45,0]$_s$, which does not have a 90° ply, thus making it more difficult to identify $E_2$ from the traction test in the 1-direction.

We also note that some of the correlation coefficients are significant. The correlation structure of the identified properties is an important result and we could not find any previous study giving the correlation matrix of the orthotropic constants identified. Ignoring the correlation could lead to significantly overestimating the uncertainty in the identified properties. The significance of this result can be illustrated by an example in probabilistic structural design (e.g. reliability analyses), which use variability models in order to estimate the probability of failure of a structure. This variability can be estimated or propagated through the physics of the problem. In all the cases an important part of the total uncertainty stems from the measurements. Uncorrelated uncertainty models are often used for the experimental uncertainty due to lack of better estimates and this can lead to errors in the probability of failure. The Bayesian identification approach offers the possibility to improve the models of experimental uncertainty by providing correlation data. Initial studies on the impact of the correlation models on experimental uncertainty were presented in [23].

At this point we also want to make a short note on the significance of the coefficients of variation and the correlation matrix. These account for the fact that for the given layup of the specimen and for the given open hole test the displacement fields have specific sensitivities with respect to each of the four ply-elastic constants. If the displacement field is highly sensitive to one of the elastic constants then this constant will be identified with a low coefficient of variation. If two elastic constants affect the displacement fields in a similar way then these two constants will be identified with some correlation. This also means that changing the experimental measurement technique but keeping the same layup and the same test this may change the values of the coefficients of variation and correlations. For our test the changes in the correlation structure turn out to be rather small when a different signal to noise ratio is used so that even with a different measurement technique a similar correlation structure is obtained. On the other hand changing the ply layup or using a different test (e.g. Iosipescu, Brazilian disc) would drastically change the sensitivities and thus the coefficients of variations and the correlation matrix. Basically the identified variance-covariance structure is specific to the test, specimen layup and uncertain parameters being considered.

The identified mean values in Table 6 show a good agreement with the manufacturer's specifications (cf. Table 1), except for $E_1$. This might seem surprising, however Noh [11] found a similar value on the same prepreg roll that we used (cf. Table 1). The mean values of $E_2$, $\nu_{12}$ and $G_{12}$ are close to the specification values. $G_{12}$ is far however from Noh's values but it should be noted that the four point bending test used by Noh is relatively poor for identifying $G_{12}$.

While it might seem surprising that the property that is identified with the lowest uncertainty ($E_1$) is also the one which is the furthest away from the manufacturer's specifications, it is important to recall that the identification does not account for inter-specimen variability or inter-prepreg batch variability of the material properties. Thus if the specimen or the prepreg roll deviates somewhat from the manufacturer's specification, it is not contradictory that, while identifying a property far away from the specifications, this can still be the property identified with the lowest uncertainty based on the particular experiment used. The other variabilities, not identified by the Bayesian method, would then have to be estimated by repeating tests on multiple specimens coming from different prepreg rolls. In addition, a key reason for carrying out experiments on complex laminates instead of unidirectional ones is the possibility that the effective material properties change somewhat due to interactions, manufacturing, and three dimensional effects. While the uncertainties on $E_2$ and $v_{12}$ that we found based on the plate with an open-hole experiment were high, we consider the possibility that further data processing of the displacement field may allow to reduce them further. We consider that a challenge that may be picked up by others, and for this reason we have included the test data as additional on-line material for the paper.

Even if no additional narrowing of the uncertainties from this test is possible, an appealing option to further lower the uncertainties would be to use a different experiment on the same specimen which would be more sensitive to $E_2$ and $v_{12}$, thus allowing to reduce the uncertainty on them. For example a vibration experiment could be carried out on the same specimen (plate with a hole), which would allow to identify again the four ply elastic constants and their uncertainty based on measurements of the natural frequencies of the plate. The authors carried out in [22] a Bayesian identification of the orthotropic ply elastic constants of a laminate on a rectangular plate (without a hole) based on the first ten natural frequencies of the plate. We found the uncertainties on the identified $E_2$ to be of the order of 5% COV and on $v_{12}$ of the order of 12% COV. The results are not directly applicable because the plate did not have a hole and the laminate layup was different. However, they show a promise that by using a vibration experiment on the same plate with a hole it might be possible to decrease the uncertainty on $E_2$ for the particular specimen in question.

## 5 Concluding Remarks

Elastic constants are usually identified by series of tests on uni-directional laminates. However, this is not possible at times, as when one needs to identify the properties of a laminate of unknown provenance. Also, the effective elastic constants of complex laminates may be somewhat different than those of unidirectional laminates due to three dimensional interaction effects. We considered in the present article the problem of orthotropic elastic constants identification based on full-field displacement measurements on a plate with a hole. Moiré interferometry was carried out during an open-hole tensile test and provided the experimental data for the identification. Bayesian identification was used in order to identify a probability distribution for the ply-elastic constants, thus characterizing the uncertainty with which the properties can be found from the given open-hole tensile on the given specimen. A numerical approach based on proper orthogonal decomposition (POD) and response surface methodology was implemented for the Bayesian identification. It is possible that the POD-based approach did not take full advantage of the copious data provided by the full-field measurements, and for that reason that data is provided for others to use (as on line material attached to the article).

We found that the four orthotropic elastic constants are not identified with the same confidence. Furthermore some properties were identified with non-negligible correlation. While the longitudinal Young's modulus was identified with the lowest standard deviation, the transverse Young's modulus and the Poisson's ratio were identified with the highest uncertainty.

The high uncertainties could also be lowered further by using additional experiments (e.g. a vibration experiment) on the same specimen and the Bayesian identification provides a natural framework for combining properties identified from different experiments based on their uncertainty structure.

The longitudinal Young's modulus was also found to be far away from the manufacturer's specifications. This was consistent however with previous test results on the same prepreg roll using traditional four point bending tests. It is indeed important to note that the distribution determined by Bayesian identification is only part of the total uncertainty present in design problems and additional variability need to be determined by repeating tests multiple times.

Finally, it is the authors opinion that providing not only numerical values for the identified properties but also an as comprehensive as possible uncertainty structure (e.g. mean values, standard deviations, correlations) is a worthwhile undertaking since it allows more accurate representation of experimental uncertainty at various subsequent design stages as well a solid basis for combining measurements and their uncertainty stemming from different experiments.

## Acknowledgements

This work was supported in part by the NASA grant NNX08AB40A. Any opinions, findings, and conclusions or recommendations expressed in this material are those of the author(s) and do not necessarily reflect the views of the National Aeronautics and Space Administration.

## Appendix 1: Proper Orthogonal decomposition

The objective of this appendix is to provide an overview of the proper orthogonal decomposition method that is used for the dimensionality reduction of the full-fields. First the theoretical foundations of the method are presented followed by some results of its application to the plate with a hole problem.

Let us consider $U^i \in \mathbb{R}^n$, which is the vector representation of a field (e.g. displacement field). Note that $n$ is usually several thousands. We seek, based on $N$ sample vectors $\{U^i\}_{i=1..N}$, a reduced dimensional representation of the fields' variations with some input parameters.

The aim of the proper orthogonal decomposition (POD) method is to construct an optimal, reduced dimensional basis for the representation of the sample vectors (also called snapshots). In the POD approach the snapshots need to have zero mean, if this is not the case the mean value needs to be subtracted.

We denote $\{\Phi_k\}_{k=1..K}$ the vectors of the orthogonal basis of the reduced dimensional representation of the snapshots. The POD method seeks to find the basis vectors $\Phi_k$ that minimize the representation error:

$$\min \frac{1}{2}\sum_{i=1}^{N}\left\| U^i - \sum_{k=1}^{K}\alpha_{i,k}\Phi_k \right\|_{L2}^2 \quad (6)$$

Because $\{\Phi_k\}_{k=1..K}$ is an orthogonal basis, the coefficients $\alpha_{i,k}$ are given by the orthogonal projection of the snapshots onto the basis vectors. As a result we have the following reduced dimensional representation $\widetilde{U}^i$ of the vectors of the snapshot set:

$$\widetilde{U}^i = \sum_{k=1}^{K} \alpha_{i,k} \Phi_k = \sum_{k=1}^{K} \left\langle U^i, \Phi_k \right\rangle \Phi_k \qquad (7)$$

The reduction in dimension is from $N$ to $K$. The truncation order $K$ needs to be selected such as to maintain a reasonably small error in the approximate representations $\widetilde{U}^i$ of $U^i$. Selecting such a $K$ is always problem specific and an error criterion is given further down.

The main advantage of the POD method is that it provides a simple procedure for constructing the basis from the samples $\{U^i\}_{i=1..N}$. The procedure guarantees that for a given truncation order we cannot find any other basis that better approximates the snapshots subspace.

The basis $\{\Phi_k\}_{k=1..K}$ is constructed using the following matrix:

$$X = \begin{pmatrix} U_1^1 & \cdots & U_1^N \\ \vdots & \ddots & \vdots \\ U_n^1 & \cdots & U_n^N \end{pmatrix} \qquad (8)$$

The vectors $\{\Phi_k\}_{k=1..K}$ are then obtained by the singular values decomposition of $X$, or equivalently by calculating the eigenvectors of the matrix $XX^T$. The singular values decomposition allows writing that:

$$X = \Phi \Sigma \Lambda^T \qquad (9)$$

where $\Phi$ is the matrix of the column vectors $\Phi_k$. The *svd()* function in Matlab was used here for the singular value decomposition.

A truncation error criterion $\varepsilon$ is then defined by the sum of the error norms as shown in Equation 8.

$$\sum_{i=1}^{N} \left\| U^i - \sum_{k=1}^{K} \alpha_{i,k} \Phi_k \right\|_{L2}^{2} \leq \varepsilon \sum_{i=1}^{N} \left\| U^i \right\|_{L2}^{2} \qquad (10)$$

where $\varepsilon = 1 - \left( \sum_{j=1}^{K} \sigma_j^2 \bigg/ \sum_{j=1}^{N} \sigma_j^2 \right)$, and $\sigma_j$ are the diagonal terms of the diagonal matrix $\Sigma$. For a derivation of this criterion and further details on POD the reader can refer to [17]. The variation of this criterion with the truncation order for the plate with a hole problem is provided in Table 3 of the main section.

The error norm truncation criterion ε, while being a global error criterion, is relatively hard to interpret physically. Furthermore the criterion is based only on the convergence of the snapshots that served for the POD basis construction. However in most cases we will want to decompose a field that is not among the snapshots, so we also want to know the convergence of the truncation error in such cases.

Accordingly we chose to construct a different error measure based on cross validation. The basic idea of cross validation is the following: if we have $N$ snapshots, instead of using them all for the POD basis construction we can use only $N$-1 snapshots and compute the error between the actual fields of the snapshot that was left out and its truncated POD decomposition. By successively changing the snapshot that is left out we can thus obtain $N$ errors. The root mean square of these $N$ errors, which we denote by $CV_{RMS}$, is then a global error criterion that can be used to assess the truncation inaccuracy.

In order to use the cross validation technique we need to define how to measure the error between two strain fields (the actual strain field and its truncated POD decomposition). We chose the maximum absolute difference between two fields. This maximum error is computed at each of the $N$ ($N$=200 here) cross validation steps and the root mean square leads to the global error criterion $CV_{RMS}$. Table 8 provides these values for different truncation orders. The relative $CV_{RMS}$ error with respect to the value of the field where the maximum error occurs is also given in Table 8.

Table 8. Cross validation $CV_{RMS}$ truncation error criterion.

|         | K              | 2              | 3              | 4              | 5              |
|---------|----------------|----------------|----------------|----------------|----------------|
| U field | $CV_{RMS}$ (mm) | $9.35 \times 10^{-6}$ | $1.05 \times 10^{-6}$ | $1.65 \times 10^{-7}$ | $7.83 \times 10^{-8}$ |
|         | $CV_{RMS}$ (%)  | $9.96 \times 10^{-2}$ | $1.13 \times 10^{-2}$ | $2.37 \times 10^{-3}$ | $9.49 \times 10^{-4}$ |
| V field | $CV_{RMS}$ (mm) | $1.00 \times 10^{-5}$ | $6.30 \times 10^{-7}$ | $3.05 \times 10^{-7}$ | $7.32 \times 10^{-8}$ |
|         | $CV_{RMS}$ (%)  | $1.10 \times 10^{-1}$ | $4.71 \times 10^{-2}$ | $3.71 \times 10^{-3}$ | $1.84 \times 10^{-3}$ |

Next we provide a graphical representation of POD results in order to have a more intuitive understanding of the modal decomposition. First an illustration of the fields obtained for a particular snapshot (snapshot 1) is shown in Figure 6, which provides an idea of the spatial variations and order of magnitude of the displacement fields. These fields were obtained with the following parameters: $E_1$=202.2 GPa, $E_2$=10.84 GPa, $v_{12}$=0.2142, $G_{12}$=4.989 GPa, $t$=0.1312 mm.

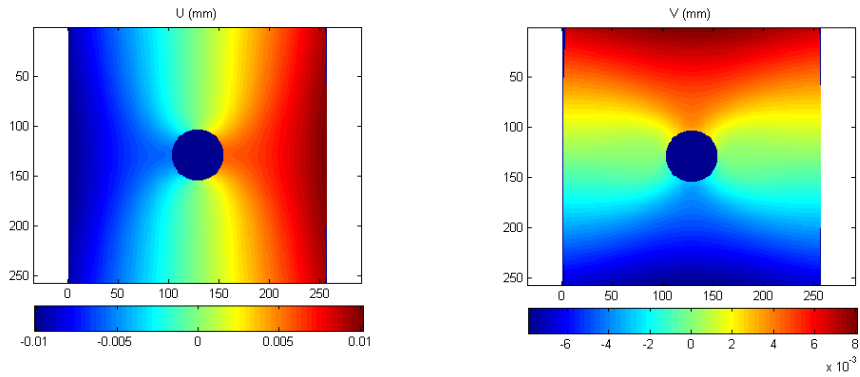

Figure 6: *U* and *V* displacement fields for snapshot 1.

The first four POD modes that we obtained are represented graphically in Figures 7 and 8. We note that the first modes are relatively close (but not identical even though the differences cannot be seen by naked eye) to the typical *U* and *V* displacement fields (see Figure 6). Furthermore we see that the modes have a more complicated shape with increasing mode number, as expected for a modal decomposition basis.

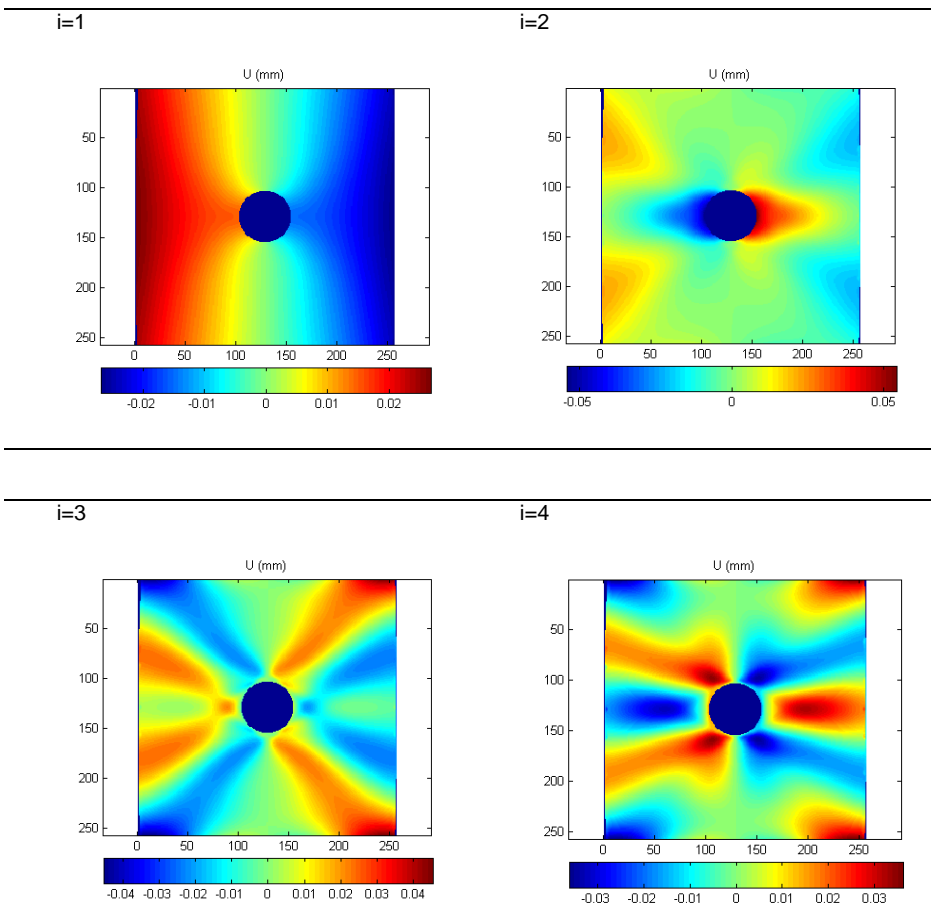

Figure 7: First 4 POD modes for the *U* field.

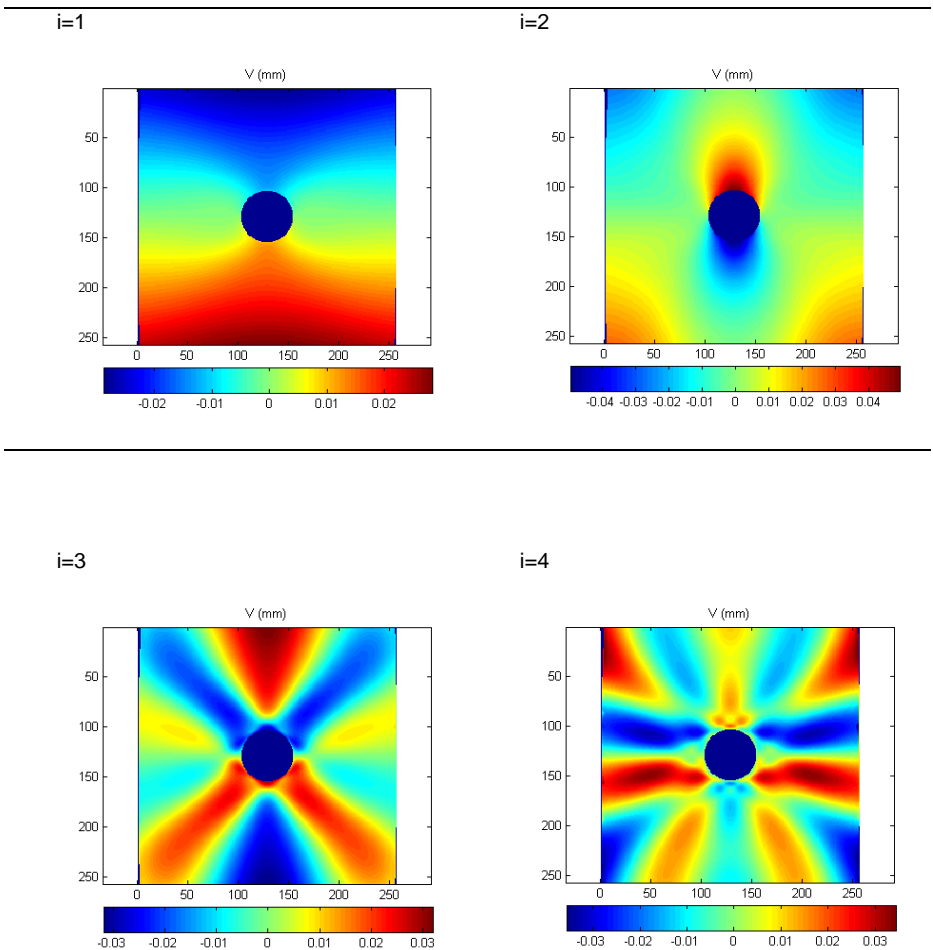

Figure 8: First 4 POD modes for the *V* field.

## Appendix 2: Error measures of the response surface approximations

The error measures used to assess the quality of the response surface approximations (RSA) constructed in section 4.2 are given in Table 9 for the first four POD coefficients of the *U* fields and in Table 10 for those of the *V* fields. The second row gives the mean value of the POD coefficient across the design of experiments (DoE). The third row provides the standard deviation of the coefficients across the DoE, which gives an idea of the magnitude of variation in the coefficients. Row four provides $R^2$, the correlation coefficient obtained for the fit, while row five gives the root mean square error among the DoE points. The final column gives the cross validation PRESS error [20].

Table 9. Error measures for RSA of the U-field POD.

| POD coefficient RSA | $\alpha_1$ | $\alpha_2$ | $\alpha_3$ | $\alpha_4$ |
|---|---|---|---|---|
| Mean value of $\alpha_i$ | $-4.04 \; 10^{-1}$ | $-3.40 \; 10^{-5}$ | $-2.20 \; 10^{-5}$ | $-8.35 \; 10^{-7}$ |
| Standard deviation of $\alpha_i$ | $8.19 \; 10^{-2}$ | $6.92 \; 10^{-4}$ | $2.01 \; 10^{-4}$ | $2.80 \; 10^{-5}$ |
| $R^2$ | 0.99999 | 0.99993 | 0.99992 | 0.99951 |
| RMS error | $2.77 \; 10^{-4}$ | $6.32 \; 10^{-6}$ | $2.01 \; 10^{-6}$ | $6.75 \; 10^{-7}$ |
| PRESS error | $3.61 \; 10^{-4}$ | $7.92 \; 10^{-6}$ | $2.67 \; 10^{-6}$ | $9.33 \; 10^{-7}$ |

Table 10. Error measures for RSA of the V-field POD.

| POD coefficient RSA | $\alpha_1$ | $\alpha_2$ | $\alpha_3$ | $\alpha_4$ |
|---|---|---|---|---|
| Mean value of $\alpha_i$ | $-2.97 \; 10^{-1}$ | $-9.51 \; 10^{-5}$ | $-2.14 \; 10^{-5}$ | $9.76 \; 10^{-7}$ |
| Standard deviation of $\alpha_i$ | $5.40 \; 10^{-2}$ | $2.26 \; 10^{-3}$ | $3.10 \; 10^{-4}$ | $1.50 \; 10^{-5}$ |
| $R^2$ | 0.99999 | 0.99992 | 0.99987 | 0.99830 |
| RMS error | $1.69 \; 10^{-4}$ | $2.26 \; 10^{-5}$ | $3.88 \; 10^{-6}$ | $6.89 \; 10^{-7}$ |
| PRESS error | $2.45 \; 10^{-4}$ | $3.05 \; 10^{-6}$ | $5.27 \; 10^{-6}$ | $1.04 \; 10^{-6}$ |

Comparing the error measures for each coefficient to their range of variation (i.e. standard deviations) we considered that the RSA are accurate enough to be used in the identification process, with the approximation error being negligible compared to the other sources of uncertainty.

## Appendix 3: Bayesian numerical implementation

A Bayesian identification numerical procedure able to account for measurement uncertainty, modeling uncertainty as well as uncertainty in other model parameters was previously developed by the authors in [21] and will also be used here.

The numerical procedure uses Monte Carlo simulations for uncertainty propagation. While the use of Monte Carlo simulation has the advantage of propagating uncertainties represented by arbitrary probability distribution functions it has the drawback of being computationally expensive. The POD method and response surface methodology are thus used for reducing the cost associated with the construction of the likelihood function. A flowchart overview of the utilized procedures is presented in Figure 9.

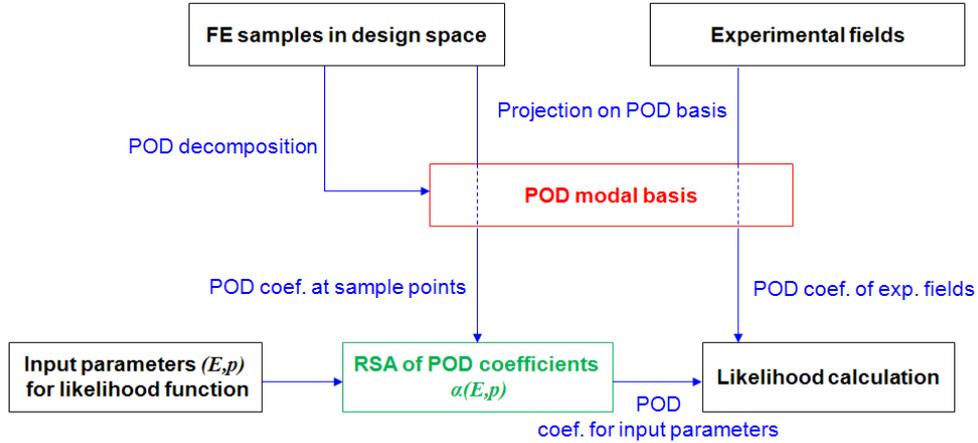

Figure 9. Flow chart of the procedure used to calculate the likelihood function. Cost reduction is shown in green and dimensionality reduction in red.

The likelihood function is computed point by point within the prior distribution's support (truncation bounds) at a grid in the four-dimensional space of the material properties $E = \{E_1, E_2, \nu_{12}, G_{12}\}$. We chose a $17^4$ grid, which is a compromise between convergence and computational cost considerations.

At each of the grid points, $E$ is fixed and we need to evaluate the probability density function (PDF) of the POD coefficients, $f_{\alpha/E=E^{fixed}}(\alpha)$, at the point $\alpha = \alpha^{measure}$. The PDF of the POD coefficients is determined by propagating through Monte Carlo with 4000 simulations the uncertainties in the other model parameters (plate thickness here) and adding a sampled value of the normally distributed uncertainty in the POD coefficients resulting from measurement and modeling uncertainty, as described in the previous subsection.

Physical considerations showed that the resulting samples must be close to Gaussian so the samples were replaced by the normal distribution, having the sample mean and variance-covariance matrix. This Gaussian nature is due to the fact that the uncertainty resulting from the measurement noise is Gaussian and the uncertainty due to thickness is proportional to $1/h$, which can in this case be well approximated by a normal distribution. The distribution $f_{\alpha/E=E^{fixed}}(\alpha)$ was then evaluated at the point $\alpha = \alpha^{measure}$, leading to $f_{\alpha/E=E^{fixed}}(\alpha^{measure})$. In this way we obtain a discretized likelihood function, which multiplied by the prior distribution gives us the posterior distribution of the elastic constants that we seek to identify.

At this point we want to make the following note. We found that the overall uncertainty on the POD coefficients is close to normal, which means that the Bayesian identification could have been treated within a purely analytical framework, thus avoiding the need for expensive Monte Carlo simulations. The analytical treatment would however have no longer been possible if uncertainties on more complex input parameters would have been considered leading to a clearly non-Gaussian distribution on the POD coefficients. In such a case the Monte Carlo simulations based approach would still work and this is the purpose why we developed this more general approach in [21]. For convenience we reused our already developed approach here, even though with several days of computational cost this approach is clearly not numerically the most efficient. More efficient numerical implementation are certainly possible and we provide the data files of the experimental results together with this paper such as to allow interested persons (including possibly ourselves at a future point) to test other implementations on the same experimental data.

## Appendix 4: Data structure of the provided measurements

The authors also provide attached to the online version of the paper the experimental displacement fields obtained. These are the displacement fields obtained based on the moiré interferometry images and the automated phase extraction procedure described in section 2.1. The data is provided in matrix form as .txt files which can be opened with any text editor. The X.txt file provides the X coordinate (in millimeters) of each point where a displacement value is provided. Similarly the Y.txt file provides the Y coordinate. The U.txt file provides the displacements in micrometers at each point in the loading direction (Y direction). The V.txt file provides the displacements in the orthogonal direction (X direction). Note that the X and Y direction are related to the actual experiment axes (Figure 2) not to the 1 and 2-driections provided in Figure 1.